\documentclass[baaa]{baaa}

\usepackage[pdftex]{hyperref}
\usepackage{subfigure}
\usepackage{natbib}
\usepackage{helvet,soul}
\usepackage[font=small]{caption}

\contriblanguage{1}

\contribtype{1}

\thematicarea{7}

\title{T Tauri stars as $\gamma$-ray source in the Rho Ophiuchi region}

\titlerunning{T Tauri stars as the $\gamma$-ray source in the Rho Ophiuchi region}

\author{A. Filócomo\inst{1,2,3}, G.J. Escobar\inst{2,3}, J.F. Albacete Colombo\inst{1}, E.A. Pássaro\inst{3} {\&} J.A. Combi\inst{2,3}}
\authorrunning{Filócomo et al.}

\contact{agostinaf@iar.unlp.edu.ar}

\institute{
{Departamento de Investigaci\'on en Ciencias Exactas e Ingenier\'ia, UNRN - Sede Atl\'antica, Viedma, Argentina}
\and
Instituto Argentino de Radioastronomía, CONICET--CICPBA, Argentina 
\and
Facultad de Ciencias Astron\'omicas y Geof\'isicas, UNLP, Argentina
}

\resumen{Más del 30{~\%} de las fuentes de {rayos $\gamma$} detectadas en el último catálogo del satélite {\sl Fermi} no tienen contraparte observacional en otras frecuencias. Una importante fracción de estas fuentes coindicen posicionalmente con regiones de formación estelar dominadas por asociaciones de estrellas T Tauri. Rho Ophiuchi, que es una de las regiones de formación estelar más cercanas, coincide con la fuente {\sl Fermi} desconocida 4FGL J1625.3-2338. En este trabajo modelamos la distribución espectral de energía considerando algunos procesos radiativos dominantes en estrellas T Tauri. Contabilizando un total de 22 estrellas T Tauri de Clase III en la región, la luminosidad integrada en {rayos $\gamma$} en el rango de energía de 100~MeV a 100~GeV es consistente con la observada en el catálogo para la fuente 4FGL J1625.3-2338.
}

\abstract{More than 30{~\%} of $\gamma$-ray sources detected in the last source catalog of the {\sl Fermi} satellite have no observational counterpart at other frequencies. A significant fraction of these sources is positionally in agreement with star-forming regions dominated by associations of T Tauri stars. Rho Ophiuchi, which is one of the closest star-forming regions, matches the unidentified {\sl Fermi} source 4FGL J1625.3-2338. In this work we modeled the spectral energy distribution considering some dominant radiative processes in T Tauri stars. Accounting for a total of 22 Class III T Tauri stars in the region, integrated $\gamma$-ray luminosity in the 100~MeV to 100~GeV energy range is consistent with the observed in the catalog for the 4FGL J1625.3-2338 source.
}

\keywords{stars: variables: T Tauri{, Herbig Ae/Be} --- gamma rays: general --- catalogs}

\begin{document}

\maketitle
\section{Introduction}
\label{intro}

The nature of unidentified $\gamma$-ray sources remains one of the most pressing problems of current high-energy astrophysics. In the last 20 years, $\gamma$-ray astronomy has reached a significant degree of maturity due to consolidation in observational technologies that can detect and measure $\gamma$-ray emission of cosmic sources, even for those without an apparent counterpart at other wavelengths.

The {Fermi Large Area Telescope Fourth Source Catalog} \citep[4FGL]{abdollahi2019}, based on eight years of continuous observations, contains more than 5\,000 $\gamma$-ray sources detected above $4~\sigma$ of confidence over the background. Many of these sources are identified with pulsars, X-ray binaries, supernova remnants, extragalactic blazars and radio galaxies \cite[e.g.][]{combi2003, combi2005}. However, a large percentage of the detected $\gamma$-ray sources in 4FGL still do not have astrophysical counterparts at lower frequencies. The first attempts to identify unknown {\sl Fermi} sources with Young Stellar Objects (YSOs) was made by \cite{munaradrover2011} through cross-correlation between the unknown sources of the {Fermi Large Area Telescope First Source Catalog} \citep{abdo2010} and a catalog of galactic young stellar clusters. They found that over  70{~\%} of the galactic unknown $\gamma$-source sample is likely associated with star-forming regions (SFRs). This opens new questions about the underlying astrophysical processes involved in the production of $\gamma$-ray emission originated in young stars.

A first theoretical approach about the emission of $\gamma$-ray radiation in T Tauri (TT) stars was presented by \cite{delvalle2011}. Here we present a preliminary computation of the spectral energy distribution (SED) of a TT star, focusing mainly on $\gamma$-ray emission mechanism via proton-proton interaction. Finally, we compare our model with a well-known population of TT stars in the Rho Ophiuchi region.

\section{Rho Ophiuchi region}
\label{rhoph}

\begin{figure*}[ht]
  \centering
  \includegraphics[width=0.99\textwidth]{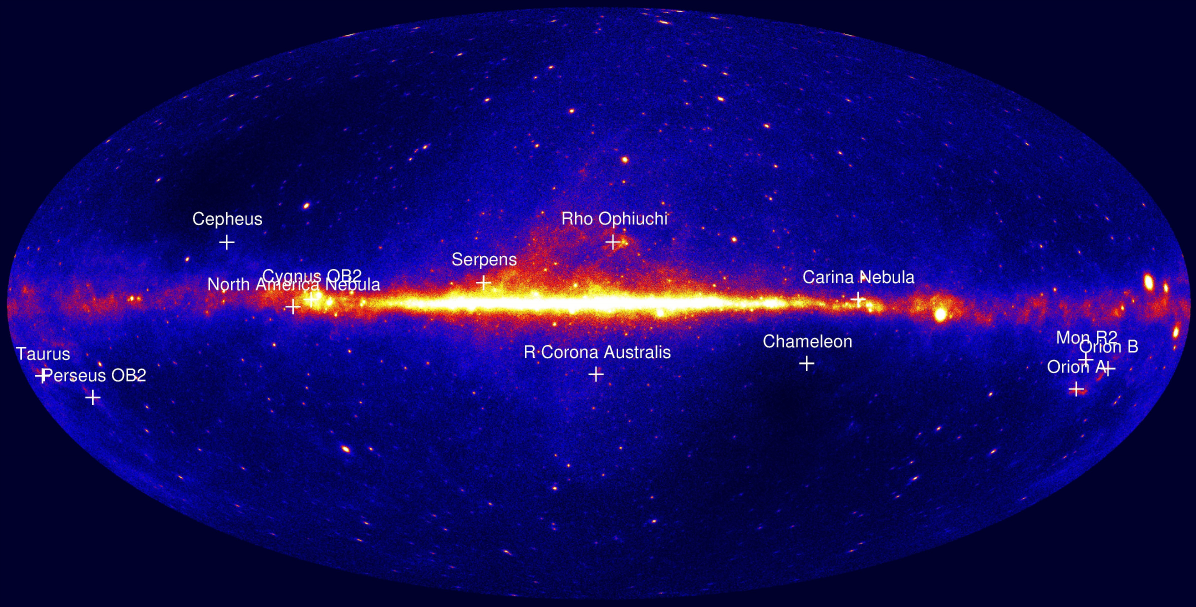}
  \caption{Aitoff projection of the 4FGL $\gamma$-ray map in the {$1-3$~GeV} energy range. White crosses indicate the center of the closest SFRs in the Galaxy.
  }
  \label{all_sky_fermi}
\end{figure*}

Rho Ophiuchi is a complex of molecular clouds with high star formation activity. It is located at {$l = 356^\circ$} and {$b = 16^\circ$} galactic coordinates (see Fig.~\ref{all_sky_fermi}) and at a very short distance of {$120 \pm 4$~pc} \citep{loinard2008}. It constitutes a very favorable constraint for multiwavelength studies, from IR \citep[e.g.,][]{santos2019}, optical \citep[e.g.,][]{wilkink2005}, and X-ray \citep[e.g.,][]{pillitteri2016}.

\cite{pillitteri2016} studied this region intending to classify 89 sources detected by the X-ray {\sl XMM-Newton} satellite. They analyzed data from {\sl XMM-Newton} and {\sl WISE} (IR) satellites. Finally, they classified 22 Class III TT stars, and 3 Transition Disks Objects (or Debris Disks, DDs). In this work, we modeled the SED for a single TT star, we scaled up its emission to the 22 sources confirmed in the region.\\

\begin{figure}[h]
  \centering
  \includegraphics[width=\columnwidth]{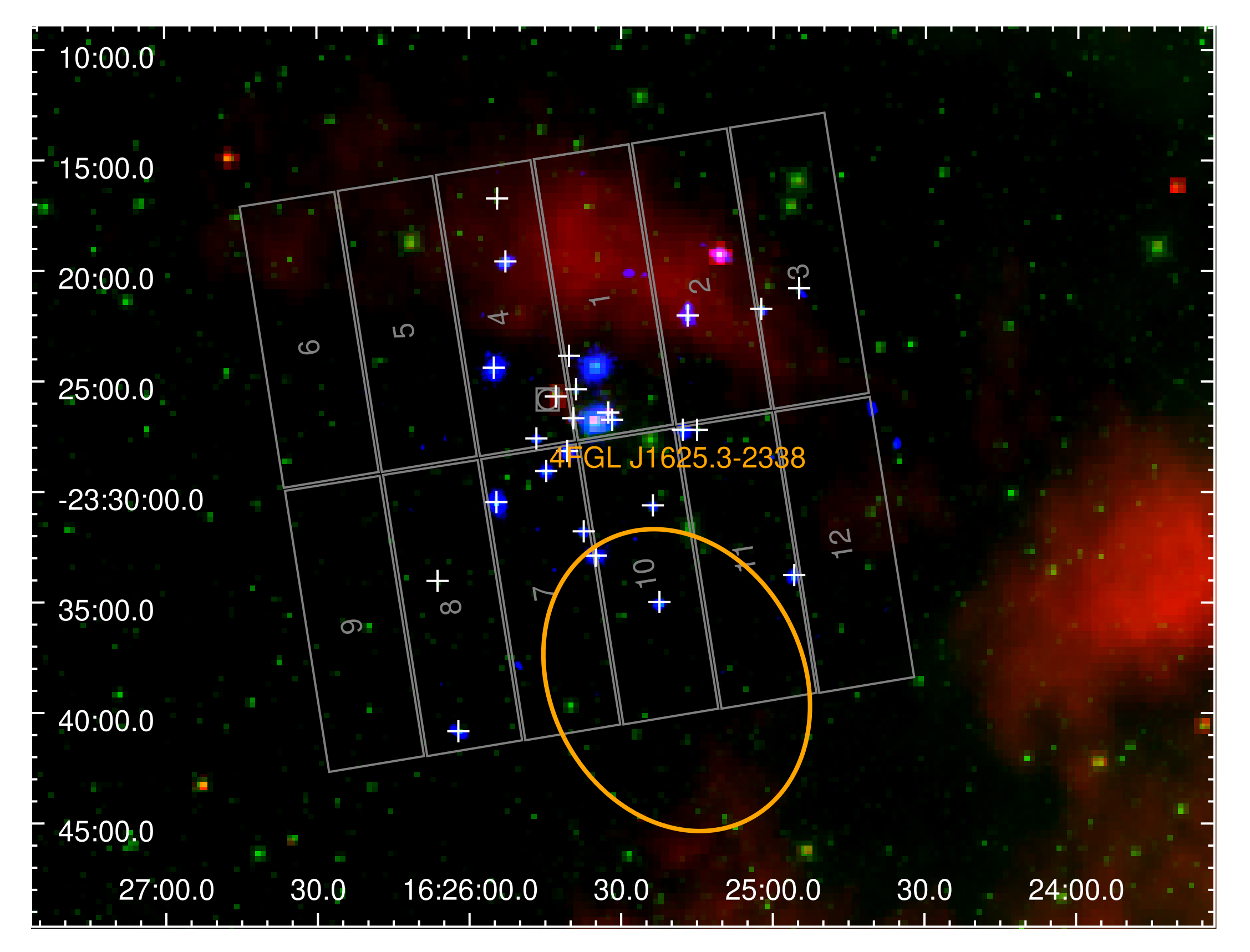}
  \caption{RGB map of Rho Ophiuchus. Red: NASA's Wide-field Infrared Survey Explorer ({\sl WISE}) mapped at 22 $\mu$. Green: {\sl WISE} mapped at 3.4 $\mu$. Blue: EPIC combined image from {\sl XMM-Newton} satellite in X-ray. White crosses indicate the position of objects analyzed by \cite{pillitteri2016}, which are YSOs classified as Class III TTs, and three of them are DDs. The field of view of the EPIC pn camera for the observation analyzed by \cite{pillitteri2016} is in grey. The orange ellipse indicates the position of the error ellipse at 95{~\%} confidence of the source 4FGL J1625.3-2338. 
  }
  \label{rho_oph}
\end{figure}

In Fig.~\ref{rho_oph}, we show a color-coded image of the central region of the Rho Ophicuchi region. Several TT-stars lie inside the error ellipse of the {\sl Fermi} source. However, we must bear in mind that sources that lie outside the error ellipse can also contribute to the total $\gamma$-ray luminosity.

\section{$\gamma$-ray emission on T Tauri stars}
\label{model}

Theoretical studies on $\gamma$-ray emission processes in young stellar objects \cite[e.g.][]{bosch-ramon2010, delvalle2011, araudo2013} suggest that young stars in nearby SFRs are potential counterparts of unidentified $\gamma$-ray sources. Nevertheless, a complete study of the SED emitted by the TTs is unknown, and the need for new and upgraded theoretical efforts is required better to understand the nature of some unidentified {\sl Fermi} sources. To discern whether the contribution of TT stars is responsible for $\gamma$-ray emission detected by the {\sl Fermi} telescope, we are developing an emission model for the SED of this kind of sources. In this work we followed the basis of the emission model given in \citet{delvalle2011}.

TT stars are mainly composed by a central star, an accretion disk, and flux tubes through which the star accretes material from the disk. In our model we assumed protons and electrons accelerate up to relativistic energies in large loops of the star through magnetic reconnection and reach the magnetosphere. In this region, these populations of relativistic particles interact with matter, radiation, and magnetic fields, producing non-thermal radiation. The energy deposited in these relativistic particle populations can be assumed as a fraction of the kinetic energy in equipartition with that of the magnetic field in the flux tubes.

We model the SED for two radiative processes: synchrotron emission for electrons and protons, and photon emission through the decay of neutral pions produced in proton-proton inelastic collisions. In order to compute the relativistic-particle densities, we must take into account the loss and acceleration mechanisms operating in these particles populations. The maximum energy values of the populations are given by equating loss and acceleration rates, where the upper limit is imposed by the Hillas criterion. The acceleration rate is given by $t_{\mathrm{acc}}^{-1} = \eta\,e\,c\,B / E$, where $\eta$ is the acceleration efficiency of the source.\footnote{In a magnetized plasma we can estimate \mbox{$\eta \approx 0.1\,r_{\mathrm{g}}\,c\,D^{-1}\left( \frac{v_{\mathrm{rec}}}{c} \right)^{2}$}, where $D$ is the diffusion coefficient, $r_{\mathrm{g}}$ the particle gyroradius, and $v_{\mathrm{rec}}$ the reconnection velocity.} To compute the synchrotron loss rate we use formulae given in \citet{Blumenthal1970}. The loss rate for proton-proton inelastic collisions is given by $t_{pp}^{-1} \approx \sigma_{pp}^{\mathrm{inel}}\,n_{\mathrm{t}}\,c$ \citep[e.g.][]{Begelman1990}, being $\sigma_{pp}^{\mathrm{inel}}$ the cross section of the process and $n_{\mathrm{t}}$ is the target-particles density. In addition, particles may lose energy through adiabatic losses and by escaping from the region of interest by convection ($t_{\mathrm{conv}}^{-1} = v_w\,l^{-1}$, where $v_w$ is the wind velocity and $l$ is the loop length). For the relativistic proton population, losses are dominated by proton-proton collisions, while for electrons, synchrotron losses are dominant. The maximum energy of each population is obtained through the condition $t_{\mathrm{acc}}^{-1} = t_{\mathrm{loss}}^{-1}$, where $t_{\mathrm{loss}}^{-1}$ accounts for all the energy-loss mechanisms.

Two important parameters are the magnetic field strength and the particle density since the first one is related to the energy available to accelerate particles. The particle density indicates the number of targets with which the accelerated particles can interact in proton-proton inelastic collisions. If we assume the medium is mainly composed by ionized Hydrogen, the density of electrons and protons are similar, then we fixed the particle density at {$n = 5 \times 10^{11}~\text{cm}^{-3}$} \citep[see][]{delvalle2011}. In order to be conservative, we assumed a magnetic field of {$B = 300$~G}, since the expected values for TTs might even rise up to $\sim 1~\text{kG}$ \citep[e.g.,][]{hill2017}.

\begin{figure}
  \includegraphics[width=\columnwidth]{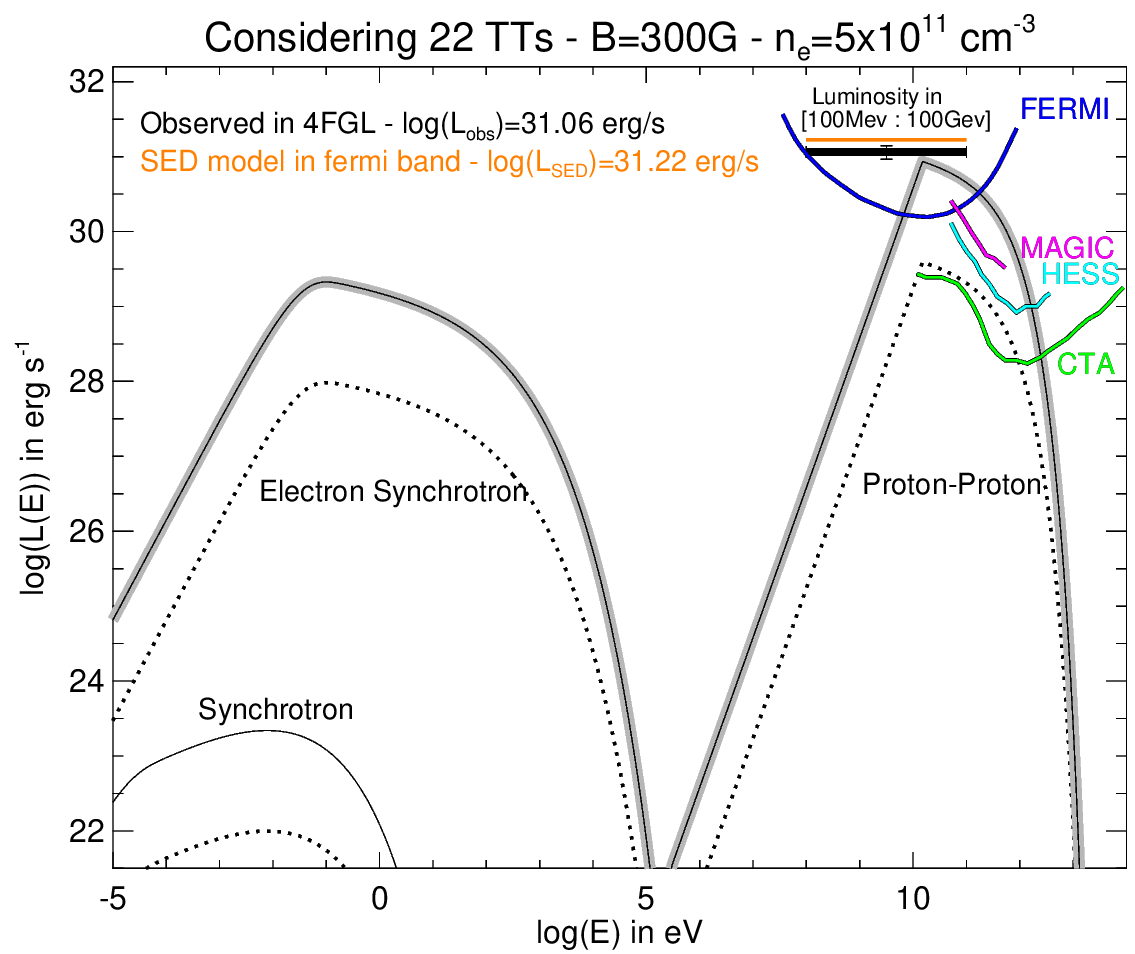}
  \caption{Computed luminosity SED for a single TT star (dotted lines) with typical magnetic field of 300~G and electron density of {$5 \times 10^{11}~\text{cm}^{-3}$}. Continuous lines refer to stacked contribution from 22 TTs in the region. The black bar is the integrated luminosity detected by {\sl Fermi} and its uncertainty, and the orange one is the integrated theoretical luminosity. The {\sl Fermi} upper sensitivity window at {$l = 0^\circ$ and $b = 16^\circ$} in the 100~MeV to 100~GeV broadband is shown in blue line. Upper sensitivity curves for Cherenkov telescopes that operate in the $\gamma$-ray range are also shown.}
  \label{tt_sed}
\end{figure}

Fig.~\ref{tt_sed} shows the computed SED for a single TT star together with the emission model accounting the whole population of such stars, which consists of at least 22 members \citep{pillitteri2016}.

The source flux in the {\sl Fermi} catalog is the integrated flux in the {$100~\text{MeV} - 100~\text{GeV}$} energy range, so we integrate the computed model luminosity in that range to be able to compare it with the observed one (see Fig.~\ref{tt_sed}).

\section{Conclusion and discussion}
\label{conclusion}

In this work, we computed the SED for a total of 22 TTs located in the Rho Ophiuchi region, which allows us to compare the $\gamma$-ray luminosity with the one detected by the {\sl Fermi} satellite. The integrated luminosity in the range of 100~MeV to 100~GeV (showed in Fig.~\ref{tt_sed}) is {$1.67\times 10^{31}~\text{erg}\,\text{s}^{-1}$}, according to the modeled SED, while that observed by {\sl Fermi} is $1.15^{(1.39)}_{(0.91)}\times10^{31}~\text{erg\,s}^{-1}$ with upper and lower luminosity values at 1~$\sigma$ uncertainty.

The integrated luminosity predicted by the model exceeds the observed one in about 40{~\%} (see Fig.~\ref{tt_sed}). We understand that our SED model is incomplete, and at the same time, the input parameters of the model would be biased or misestimated. Forthcoming improvements to our model focus on the p-$\gamma$ interaction. We estimate that protons accelerated in the flux tubes interact with X-ray photons, probably contributing to $\gamma$-ray luminosity. The modeling of the absorption due to photon annihilation would also strongly modify the SED. Both improvements positively contribute substantially to the study of the $\gamma$-ray emission of TTs, which remains unclear today. In future work we will analyze how the SED varies as different stellar parameters are modified.

\begin{acknowledgement}
We thank all LOC and SOC members for the successful meeting held in September 2019, Viedma, Argentina.  AF is PhD fellow and JFAC is a staff researcher of the CONICET. Both acknowledges support from UNRN (PI 40-C-691).
\end{acknowledgement}


\bibliographystyle{baaa}
\small
\bibliography{bibliografia}

\begin{thebibliography}{15}
\providecommand{\natexlab}[1]{#1}

\bibitem[{{Abdo} et~al.(2010)}]{abdo2010}
{Abdo} A.A., et~al., 2010, \apjs, 188, 405

\bibitem[{{Abdollahi} et~al.(2019)}]{abdollahi2019}
{Abdollahi} S., et~al., 2019, in press

\bibitem[{{Araudo} et~al.(2013){Araudo}, {Bosch-Ramon} \&
  {Romero}}]{araudo2013}
{Araudo} A.T., {Bosch-Ramon} V., {Romero} G.E., 2013, \mnras, 436, 3626

\bibitem[{{Begelman} et~al.(1990){Begelman}, {Rudak} \&
  {Sikora}}]{Begelman1990}
{Begelman} M.C., {Rudak} B., {Sikora} M., 1990, \apj, 362, 38

\bibitem[{{Blumenthal} \& {Gould}(1970)}]{Blumenthal1970}
{Blumenthal} G.R., {Gould} R.J., 1970, Reviews of Modern Physics, 42, 237

\bibitem[{{Bosch-Ramon} et~al.(2010)}]{bosch-ramon2010}
{Bosch-Ramon} V., et~al., 2010, \aap, 511, A8

\bibitem[{{Combi} et~al.(2005){Combi}, {Rib{\'o}} \& {Mirabel}}]{combi2005}
{Combi} J.A., {Rib{\'o}} M., {Mirabel} I.F., 2005, \apss, 297, 385

\bibitem[{{Combi} et~al.(2003)}]{combi2003}
{Combi} J.A., et~al., 2003, \apj, 588, 731

\bibitem[{{del Valle} et~al.(2011)}]{delvalle2011}
{del Valle} M.V., et~al., 2011, \apj, 738, 115

\bibitem[{{Hill} \& {MaTYSSE Collaboration}(2017)}]{hill2017}
{Hill} C., {MaTYSSE Collaboration}, 2017, \textit{IAU Symposium}, vol. 328,
  101--106

\bibitem[{{Loinard} et~al.(2008)}]{loinard2008}
{Loinard} L., et~al., 2008, \apjl, 675, L29

\bibitem[{{Munar Adrover, et al.}(2011)}]{munaradrover2011}
{Munar Adrover, et al.}, 2011, \textit{Highlights Astroph.VI}, 543--543

\bibitem[{{Pillitteri} et~al.(2016)}]{pillitteri2016}
{Pillitteri} I., et~al., 2016, \aap, 592, A88

\bibitem[{{Santos} et~al.(2019)}]{santos2019}
{Santos} F.P., et~al., 2019, \apj, 882, 113

\bibitem[{{Wilking} et~al.(2005)}]{wilkink2005}
{Wilking} B.A., et~al., 2005, \aj, 130, 1733

\end{thebibliography}
 
\end{document}